\documentclass[12pt]{article}
\usepackage{graphicx, amsmath, amssymb, cite, setspace, color, relsize, bm, bbold, array, hyperref, dsfont, bbold, xcolor, cancel,soul,placeins,hyperref,amsfonts,pifont}

\usepackage[top=1 in, bottom=1 in, left=.9 in, right=.9 in]{geometry}

\newcommand\To{\rule{0pt}{4.5ex}}       
\newcommand\Bo{\rule[-3.0ex]{0pt}{0pt}} 

\newcommand{\captionfonts}{\footnotesize} 
\makeatletter
\long\def\@makecaption#1#2{%
  \vskip\abovecaptionskip
  \sbox\@tempboxa{{\captionfonts #1: #2}}%
  \ifdim \wd\@tempboxa >\hsize
    {\captionfonts #1: #2\par}
  \else
    \hbox to\hsize{\hfil\box\@tempboxa\hfil}%
  \fi
  \vskip\belowcaptionskip}
\makeatother

\def\lsim{ \lower .75ex \hbox{$\sim$} \llap{\raise .27ex
\hbox{$<$}} }
\def\gsim{ \lower .75ex \hbox{$\sim$} \llap{\raise .27ex
\hbox{$>$}} }

\renewcommand{\title}[1]{\vbox{\center\LARGE{#1}}\vspace{5mm}}
\renewcommand{\author}[1]{\vbox{\center#1}\vspace{5mm}}
\newcommand{\address}[1]{\vbox{\center\em#1}}

\newcommand{\be}{\begin{equation}}
\newcommand{\bea}{\begin{eqnarray}}
\newcommand{\eea}{\end{eqnarray}}
\newcommand{\beq}{\begin{equation}}
\newcommand{\ee}{\end{equation}}

\let\oldsqrt\sqrt
\def\sqrt{\mathpalette\DHLhksqrt}
\def\DHLhksqrt#1#2{%
\setbox0=\hbox{$#1\oldsqrt{#2\,}$}\dimen0=\ht0
\advance\dimen0-0.2\ht0
\setbox2=\hbox{\vrule height\ht0 depth -\dimen0}%
{\box0\lower0.4pt\box2}}

\begin{document}
\begin{titlepage}

\rightline{}
\bigskip
\bigskip\bigskip\bigskip\bigskip
\bigskip

\title{Playing Pool with $|\psi \rangle$: \\
from Bouncing Billiards to Quantum Search}

\bigskip
\begin{center}

\author{Adam R. Brown}

\address{Google, Mountain View, CA 94043, USA}

\address{Department of Physics, Stanford University, Stanford, CA 94305, USA}

\end{center}

\begin{center}
\bf     \rm

\bigskip

\end{center}

\begin{abstract}

\noindent In ``Playing Pool with $\pi$'' \cite{Galperin}, Galperin invented an extraordinary method to learn the digits of $\pi$ by counting the collisions of billiard balls. Here I demonstrate an exact isomorphism between Galperin's bouncing billiards and Grover's algorithm for quantum search. 
This provides an illuminating way to visualize Grover's algorithm.

\end{abstract}

\hspace{2cm}

\let\thefootnote\relax\footnotetext{email: \tt{mr.adam.brown@gmail.com}}

\end{titlepage}





\section{Overview} \label{sec:overview}
\noindent 
An impractical but picturesque way to determine the digits of $\pi$ is to  hurl a heavy ball towards a light ball that has its back to  a wall, as in Fig.~\ref{fig-bouncing1}, and then count the ensuing elastic collisions. 
     \begin{figure}[htbp] 
    \centering
    \includegraphics[width=.3\textwidth]{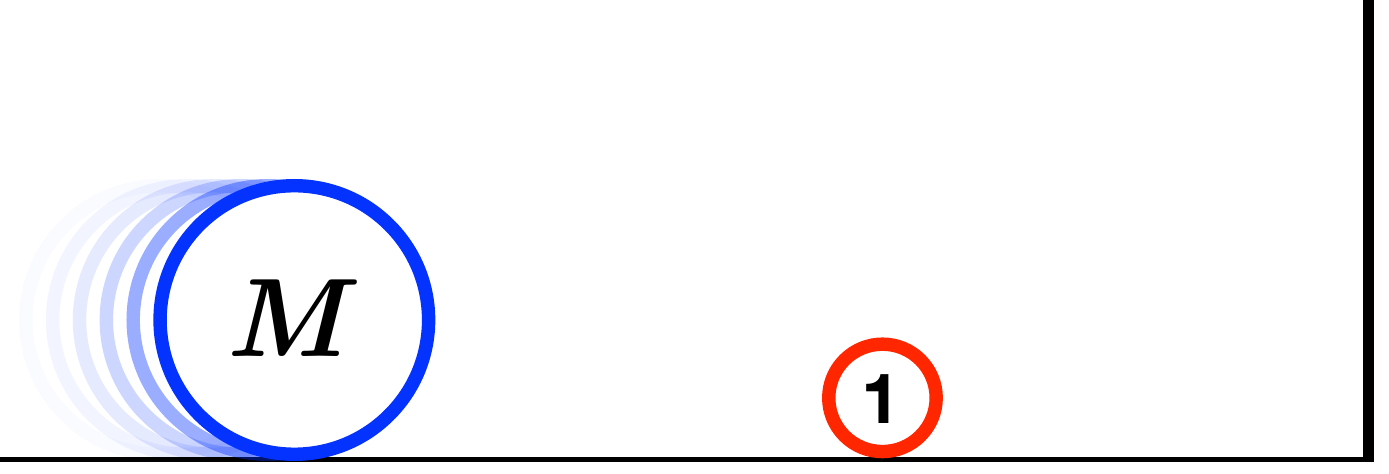} 
        \label{fig-bouncing1}
    \caption{A {\textcolor{blue}{heavy ball}}  of mass $M$ approaches a stationary {\textcolor{red}{light ball}} of mass 1. How many  collisions ensue?}
 \end{figure}

\noindent Let's start with equal masses, $M$=1. At the first collision, the left ball transfers all its momentum to the right ball. At the second collision, the right ball's momentum is reversed by the wall. At the third and final collision, the right ball transfers all its momentum back to the left ball. In total, 
\begin{equation}
M= 1 \ \rightarrow \ \#_\textrm{collisions} = 3 \ . 
\end{equation}
\vspace{-8mm}     \begin{figure}[htbp] 
    \centering
    \includegraphics[width=\textwidth]{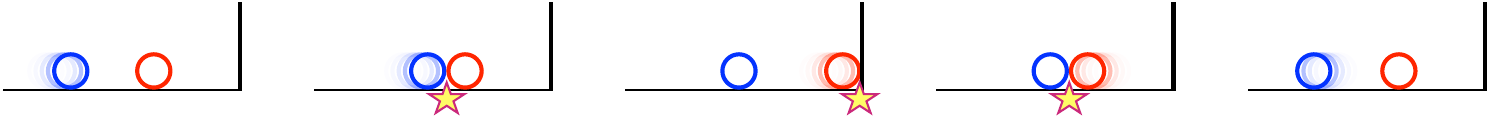} 
    \label{fig-bouncing1}
 \end{figure}

\noindent If the left ball is much heavier than the right, then it is harder to slow down and reverse. 
 The heavier the left ball, the more collisions are needed, 
\begin{eqnarray}
M = 100  \hspace{2pt} & \rightarrow & \#_\textrm{collisions} = 31\\
M = 10^6  \hspace{4pt} & \rightarrow & \#_\textrm{collisions} = 3141 \\
M = 10^{20}   & \rightarrow & \#_\textrm{collisions} = 3 1415926535 
\end{eqnarray}
These digits look familiar! In ``Playing Pool with $\pi$'' \cite{Galperin}, G.~Galperin proved that this algorithm really is spitting out the digits of $\pi$, since  for $M = 100^N$
\begin{equation}
 \#_\textrm{collisions} = \Bigl\lfloor \pi   \sqrt{M } \Bigl\rfloor \ . \label{eq:numbercollisions}
\end{equation}
I recommend reading the very readable Ref.~\cite{Galperin} and watching the very watchable Ref.~\cite{youtube}. 

\vspace{5mm}

\noindent Let us make a seemingly abrupt shift and now turn our attention to the field of quantum query complexity. One of the most famous  algorithms in all of quantum mechanics  is that of L.~Grover \cite{Grover:1996rk}. Grover's algorithm provides a way to  ``find a needle in a quantum haystack''---or more precisely to determine which out of $d$ mystery functions are being implemented by a black-box quantum oracle. For $d -1= 100^N$ the runtime is 
\begin{equation}
\textrm{number of oracle calls} =  \Bigl\lfloor \frac{1}{4} \pi \sqrt{d-1} \Bigl\rfloor . \label{eq:numbergrovercalls}
\end{equation}
In this paper I will argue 
that the similarity between Eq.~\ref{eq:numbercollisions} and Eq.~\ref{eq:numbergrovercalls} is not a coincidence. It's the same squareroot, and the same $\pi$, and for the same reason. The factor of $\frac{1}{4}$ is merely a book-keeping anomaly arising from different accounting practices. Indeed, I will argue that 
there is a precise isomorphism between bouncing billiard balls and quantum search.

\section{Quantum Search}
The wavefunction of a $d$-state system---a `qudit'---may  in general be written as
\begin{equation}
|\psi \rangle = v_1 |1 \rangle +  v_2 |2 \rangle + \ldots +  v_{d-1} |d-1 \rangle  +  v_d |d \rangle \, ,
\end{equation}
and conservation of probability means that throughout its evolution the state will always have
\begin{equation}
\textrm{conservation of probability:} \ \  \ \ \ \ \ \langle \psi | \psi \rangle = \sum_{i = 1}^d |v_i|^2 = 1 \ . \hspace{5cm}
\end{equation}
The Grover task \cite{Grover:1996rk} imagines we are given a black box that implements the transformation 
\begin{equation}
\hat{U}_w \equiv \mathds{1}  - 2 |w \rangle \langle w | \ .
\end{equation}
This unitary acts on the qudit by flipping the sign of the $w^\textrm{th}$ amplitude while leaving all the others unchanged. For example, for $w=7$ we'd have
\begin{equation}
\hat{U}_7 | \psi \rangle = v_1 |1 \rangle + \ldots + v_{6} |6 \rangle - v_7 |7\rangle + v_{8} |8 \rangle + \ldots + v_d |d \rangle. 
\end{equation}
 Our job is to figure out which $\hat{U}_w$ we have been given, i.e.~to determine the value of $w$. Making our task harder is the fact that the value of $w$  is not written on the box, and we are not allowed to take a screwdriver to open the box: the rules state  that the only way we are allowed to interact with the black box is to feed states in and see what comes out. \\
 
A simple strategy to determine $w$ is to test each possibility in turn, first $|1 \rangle$, then $| 2 \rangle$, then $|3 \rangle$, etc., until we come to a state whose sign is flipped. While this strategy succeeds, it succeeds only slowly---on average we will need to use the box $\frac{1}{2} d$ times. That is how long it takes to find a classical needle in a classical haystack. The surprising fact that Grover discovered is that we can do better---much better. Grover's algorithm allows us to find $w$ using only O($\sqrt{d}$) calls. 

Given that we do not know the value of $w$, the most democratic option is to start with an even superposition, and indeed that is where Grover's algorithm begins, 
\begin{equation}
| s \rangle = \frac{1}{\sqrt{d}} \Bigl( |1 \rangle + |2 \rangle + \ldots + | d -1 \rangle + |d \rangle \Bigl).
\end{equation}
We feed this into the black box, yielding $\hat{U}_w |s \rangle$, which has $- v_w = v_{i \neq w} = 1/\sqrt{d}$. Already this wavefunction `knows' the value of $w$, since $\hat{U}_w | s \rangle \neq \hat{U}_{w'} | s \rangle$ for $w \neq w'$, and so if we could determine the wavefunction directly we could solve the Grover problem in one step. But quantum mechanics doesn't work that way. The dramatic tension at the heart of quantum information theory is the interplay between the quantum `superpower'---the ability to try all possibilities at once---and the quantum `superweakness'---the limitation to always acting linearly. One consequence of linearity is that only orthogonal states can be reliably distinguished. Since for large $d$ the states $\hat{U}_w |s \rangle$ and $\hat{U}_{w'} |{s} \rangle$ are far from orthogonal, in order to determine the value of $w$ we must amplify the difference. But how? The most obvious move is just to plug the output back into the black box again, but this is counterproductive since it takes us back to square one, $\hat{U}_w^{\, 2} = \mathds{1}$. Instead, Grover showed that our next step should be to act with 
\begin{equation}
\hat{U}_s = 2 |s \rangle \langle s | - \mathds{1}. 
\end{equation}
We can construct $\hat{U}_s$ without using the black box, since it makes no reference to $w$ and indeed treats all basis states the same. Grover's algorithm is then just to repeatedly iterate $\hat{U}_s$ and $\hat{U}_w$. \\

  \begin{figure}[htbp] 
    \centering
    \includegraphics[height=2.2in]{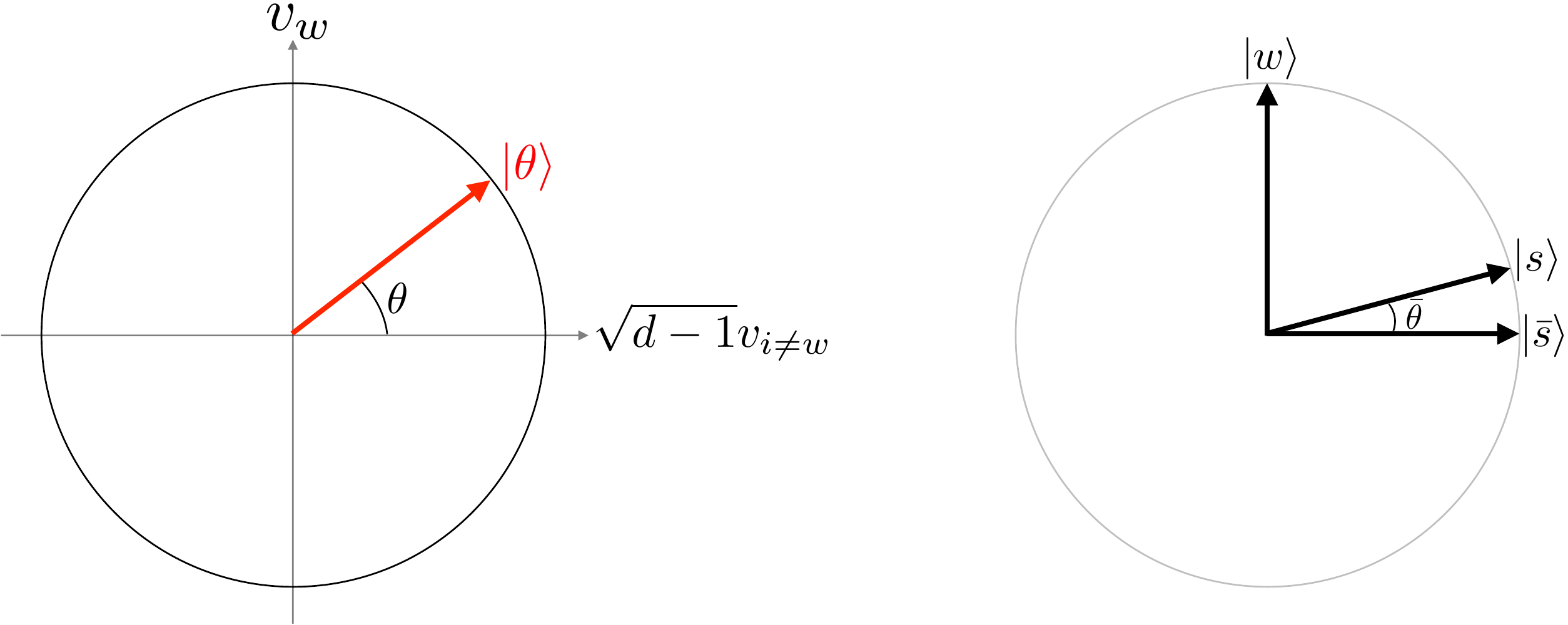} 
    \caption{All operations keep us on the circle defined by $| \theta \rangle \equiv  \cos \theta |\bar{s} \rangle + \sin \theta  |w \rangle$. The starting state $|s \rangle$ and the target state $| w \rangle$ are not exactly orthogonal, but we can define another state $|\bar{s} \rangle$ such that $\langle \bar{s} | w \rangle = 0$. }
    \label{fig-onthecircle}
 \end{figure}
In order to analyze the effect of this iteration, it will be helpful to use an orthonormal coordinate system. The state $|s\rangle$ is not exactly orthogonal to  $| w \rangle$, 
\begin{equation}
\sin \bar{\theta} \equiv \langle s | w \rangle = \frac{1}{\sqrt{d}},
\end{equation} 
but we can find a state $| \bar{s} \rangle$ such that  $\langle w | \bar{s} \rangle = 0$ by defining 
\begin{eqnarray}
 | \bar{s} \rangle &= &\frac{1}{\sqrt{d-1}} \sum_{i \neq w}  |i \rangle \ .
\end{eqnarray}
Both $\hat{U}_s$ and $\hat{U}_w$ keep us on the circle defined by 
\begin{equation}
| \theta \rangle \equiv  \cos \theta |\bar{s} \rangle + \sin \theta  |w \rangle.
\end{equation}
On this circle, $\hat{U}_s$ reflects about the $|s \rangle$-axis and $\hat{U}_w$ reflects about the $|\bar{s} \rangle$-axis, 
\begin{equation}
\hat{U}_w | \theta \rangle = \left( \mathds{1} - 2 |w \rangle \langle w | \right) | \theta \rangle  = \left( 2 | \bar{s} \rangle \langle \bar{s} | - \mathds{1} \right) | \theta \rangle \ . 
\end{equation}
The two consecutive non-parallel reflections combine to give a rotation by $2 \bar{\theta}$, 
\begin{equation}
\hat{U}_s \hat{U}_w | \theta \rangle = | \theta + 2 \bar{\theta} \rangle. \label{eq:multiplytorotation}
\end{equation}
We will use repeated applications of  $\hat{U}_s \hat{U}_w$ to rotate the state from $|s \rangle$ (i.e.~$\theta = \bar{\theta}$) to  $|w \rangle$ (i.e.~$\theta = \frac{\pi}{2}$) in steps of size $2 \bar{\theta}$. For almost all $d$, and in particular for all $d-1 = 100^N$, the integer closest to $(\frac{\pi}{2} - \bar{\theta})/2 \bar{\theta}$ is $\bigl\lfloor \frac{\pi}{4} \sqrt{d-1}\bigl\rfloor $, and so Grover's algorithm calls for that many steps
\begin{equation}
\left( \hat{U}_s \hat{U}_w \right)^{\lfloor \frac{\pi}{4} \sqrt{d-1}\rfloor } | s \rangle =  |  \bar{\theta} + 2  \lfloor \frac{\pi}{4} \sqrt{d-1} \rfloor \bar{\theta} \rangle =   | w \rangle + \textrm{O}\Bigl( \frac{1}{\sqrt{d}} \Bigl).
\end{equation}
Since $\langle w | w' \rangle = 0$ for $w \neq w'$, with high probability we can now just measure the qudit to learn the value of $w$ and successfully complete the Grover task. 

  \begin{figure}[htbp] 
    \centering
    \includegraphics[height=2in]{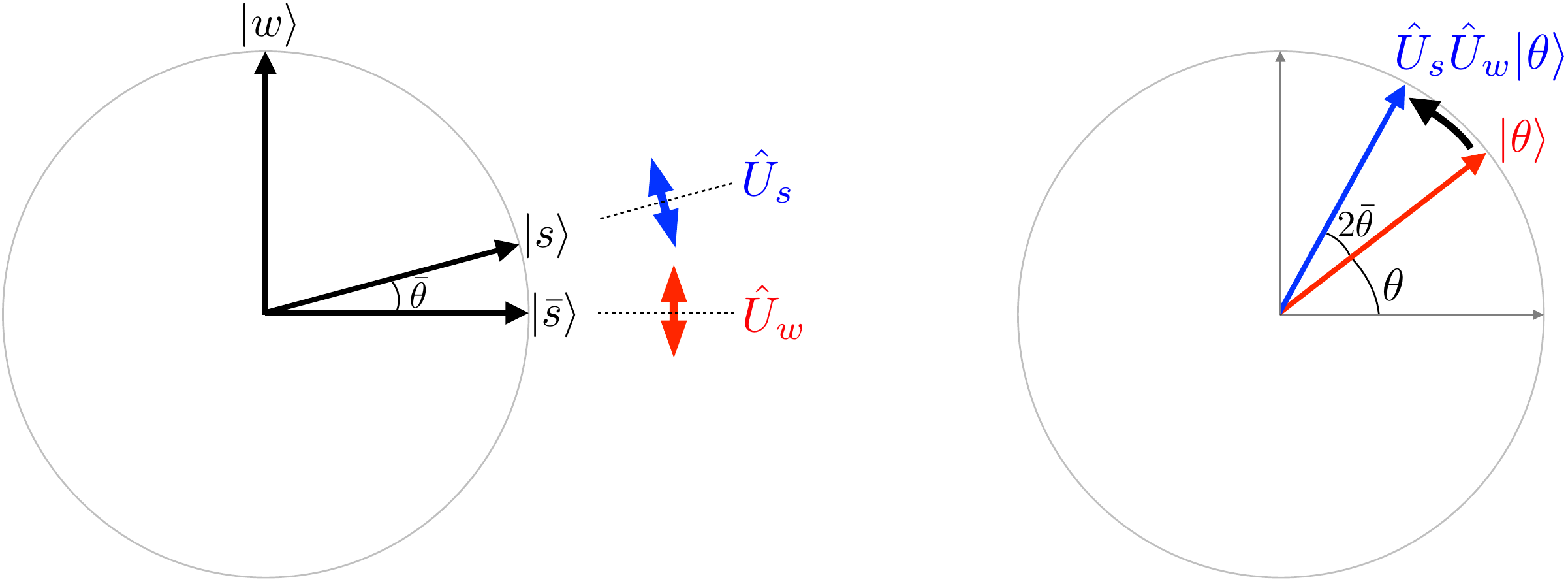} 
    \caption{$\hat{U}_w$ reflects about $|\bar{s} \rangle$, and $\hat{U}_s$ reflects about $| s \rangle$. In combination, $\hat{U}_s \hat{U}_w$ gives a rotation by $2 \bar{\theta}$. }
    \label{fig-EffectOfUwUs4}
 \end{figure}

\section{Balls to the wall} 
The velocity of $d$ billiards moving on a line may be described by a $d$-dimensional vector, $v_i$. \linebreak To make the analogy with quantum mechanics more visceral, we could write this vector in $|\textrm{ket} \rangle$ notation as 
\begin{equation}
|\textrm{velocity} \rangle = v_1 |1 \rangle +  v_2 |2 \rangle + \ldots +  v_{d-1} |d-1 \rangle  +  v_d |d \rangle.
\end{equation}
Elastic collisions conserve kinetic energy. If all the billiards have mass 1, and if they start with total kinetic energy $\frac{1}{2}$, then throughout their evolution they will preserve
\begin{equation}
\textrm{conservation of kinetic energy:} \ \ \ \ \  \ \langle \textrm{velocity} | \textrm{velocity} \rangle =  \sum_{i = 1}^d |v_i|^2 = 1 \ . \ \ \   
\end{equation}

To replicate the $\pi$-calculating set-up from Sec.~\ref{sec:overview}, we should separate off one billiard (let us say the  $w$th) to be the light ball, and glue together all the other billiards to form one big heavy ball of mass 
\begin{equation}
M= d-1 \ . 
\end{equation} 
The glue constrains all the $v_{i \neq w}$ to be the same, and then conservation of energy confines the velocity vector to lie on the circle given by $\sin \theta = v_w$ and $\cos \theta = \sqrt{M} v_{i \neq w} = \sqrt{d-1} v_{i \neq w}$,
\begin{equation}
| \theta \rangle \equiv  \cos \theta |\bar{s} \rangle + \sin \theta  |w \rangle.
\end{equation}

  \begin{figure}[htbp] 
    \centering
\ \  \ \ \        \includegraphics[height=2in]{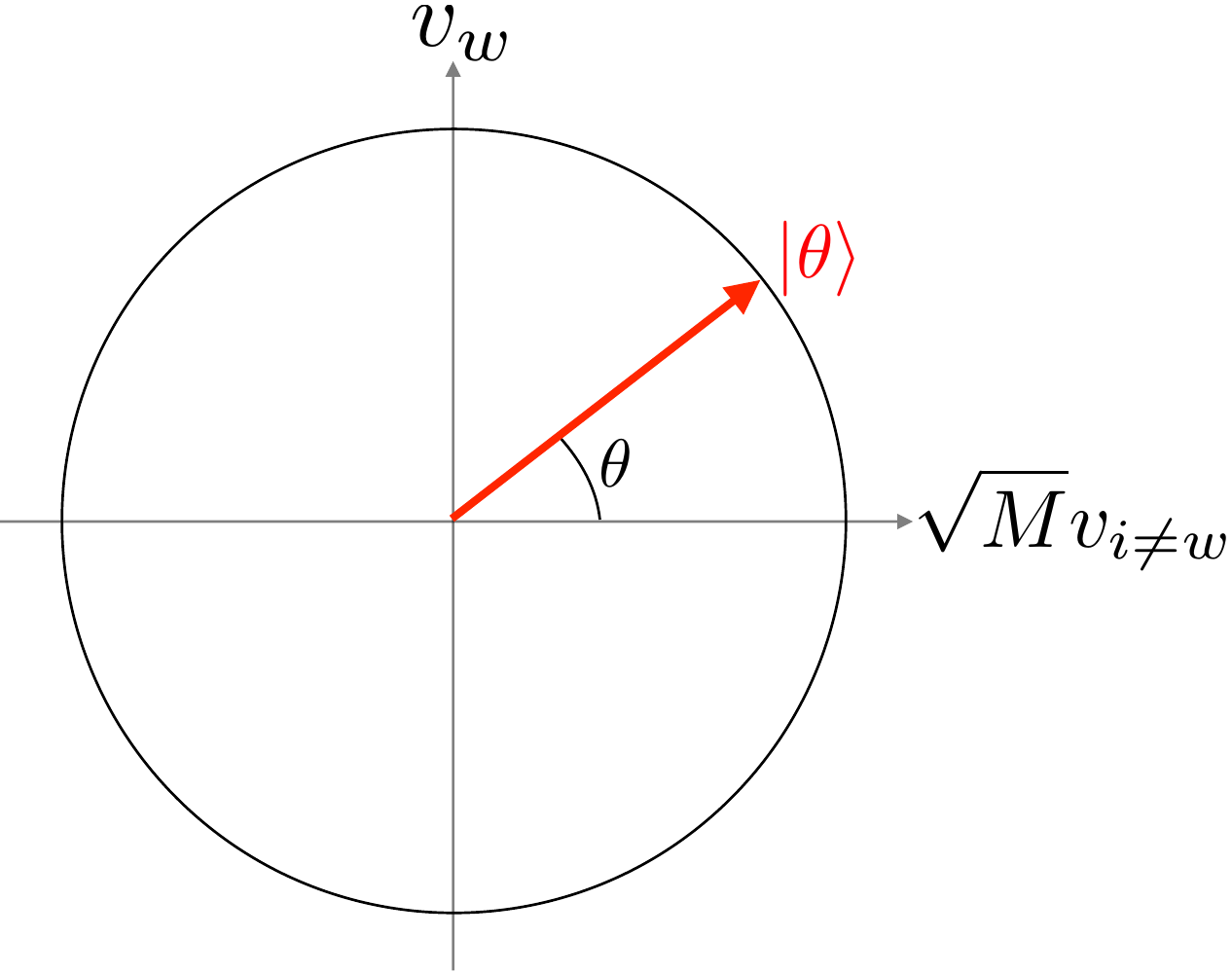} \ \ \ \ \ \ \ \ \ \ \ \ 
    \includegraphics[height=2.18in]{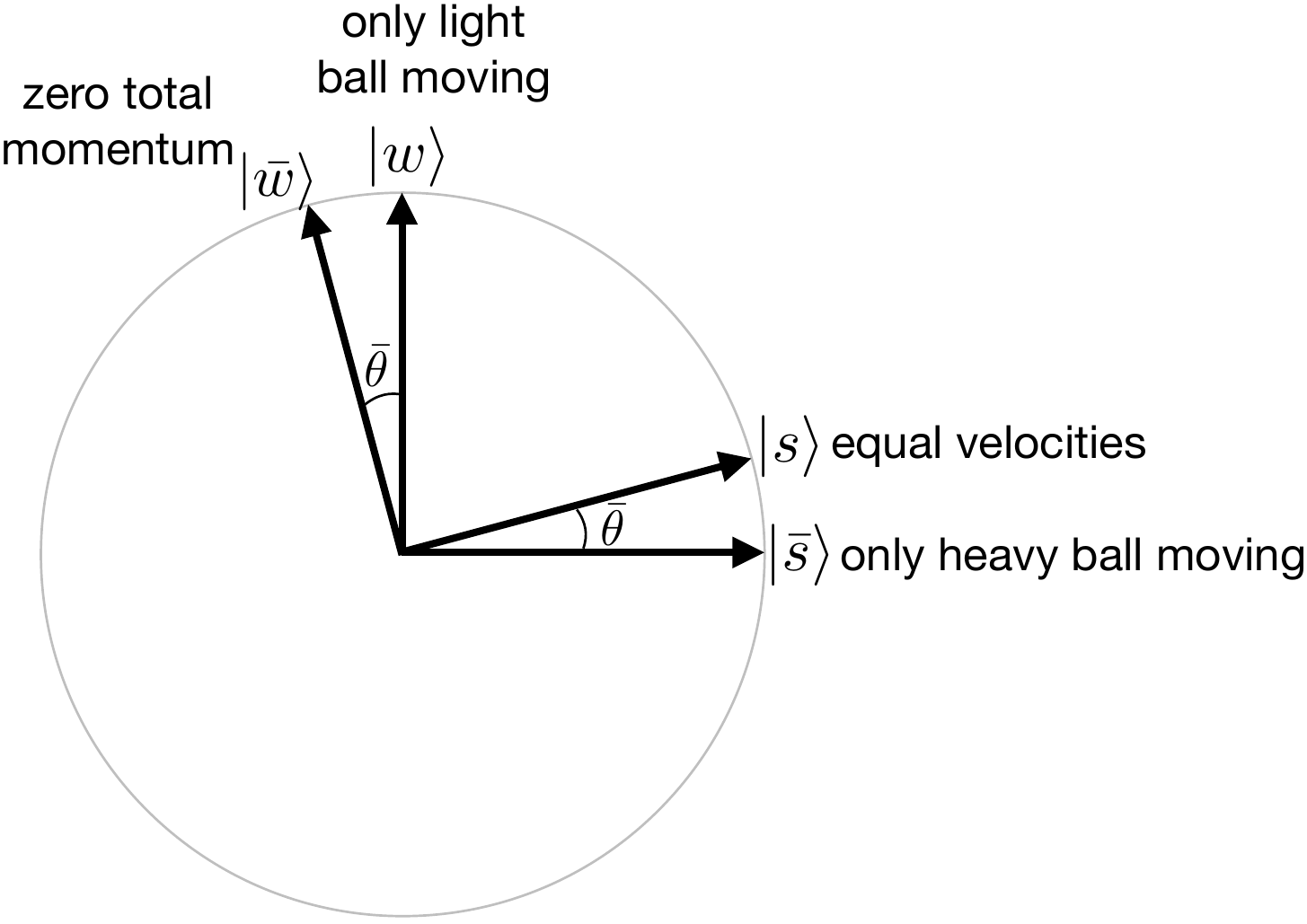}  
    \caption{Conservation of energy guarantees that the velocities are stuck on the circle  $\frac{1}{2} v_{w}^2 + \frac{1}{2} M v_{i \neq w}^2 = \frac{1}{2}$.}
    \label{fig-statedictionary}
     \end{figure}
 \begin{center}
\begin{tabular}{c||c} 
BOUNCING BILLIARDS \  \  & \ \     GROVER SEARCH  \Bo \\
\hline
\hline
all  kinetic energy in light ball & \To \Bo   $| w \rangle$ \\ 
\hline
both balls equal velocity  & \Bo   \To  $ | {s} \rangle = \frac{1}{\sqrt{d}} \sum_i  |i \rangle $ \\ 
\hline
all  kinetic energy in heavy ball  &    \To \Bo  $ | \bar{s} \rangle = \frac{1}{\sqrt{d-1}} \sum_{i \neq w}  |i \rangle $ \\ 
\hline
total momentum zero  & \Bo   \To  $ | \bar{w} \rangle = \sqrt{\frac{d-1}{d} } | w \rangle - \frac{1}{ \sqrt{d(d-1)}}  \sum_{i \neq w}  |i \rangle $  \\ 
\hline
the $M = d-1$ billiards in big ball & the $d-1$ wrong answers \Bo \To \\
\hline 
$\hat{O}_\textrm{ball}$ \ =  \ \ \  \hspace{-6pt} \raisebox{-.4\totalheight}{    \includegraphics[height=.43in]{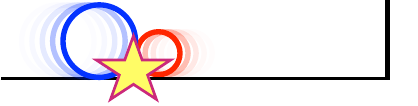} } \ \ 
  & $\hat{U}_s = 2  |s \rangle \langle s | - \mathds{1} $ \To  \Bo \\ 
\hline
$\hat{O}_\textrm{wall}$ \  =  \hspace{-1pt} \raisebox{-.4\totalheight}{    \includegraphics[height=.441in]{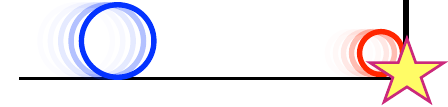} } \, & $\hat{U}_w = \mathds{1} - 2 |w \rangle \langle w |$   \To \Bo \\
\hline
small ball bounces back and forth & \To \Bo    alternate  $\hat{U}_s$ and $\hat{U}_w$ \\ 
\hline
velocity $v_i$ of $i$th  billiard &amplitude $v_i$ of $i$th eigenstate    \To \Bo \\
\hline
$2 \,  \times$ kinetic energy of $i$th  billiard & probability $|v_i|^2$ of $i$th eigenstate   \To \Bo \\
\hline
conservation of kinetic energy & conservation of probability   \To \Bo \\
\hline
conservation of phase space & unitarity   \To \Bo \\
\hline
motion purely horizontal & wavefunction purely real   \To \Bo \\
\hline
collision order matters & operators don't commute \To \Bo \\
\hline
$\hat{O}_\textrm{ball}$ conserves total momentum & $\bigl[ |s \rangle \langle s | , \hat{U}_s \bigl] = 0 $ \To \Bo \\
\hline
$\hat{O}_\textrm{wall}$ conserves big-ball momentum & $\bigl[ |\bar{s} \rangle \langle \bar{s} | , \hat{U}_w \bigl] = 0 $ \To \Bo 
\end{tabular}
\end{center}
The effect of a collision will be to map this circle into itself. The exact form of the map can be determined by considering which momentum each collision conserves. Let's argue by symmetry. The linear maps from a circle to itself are described by the group O(2), which has two components: the rotations (det = +1); and the  
reflections (det $= -1$). Rotations leave no vectors invariant, whereas reflections leave invariant the vector that points down the axis being reflected about. Thus any O(2) map that conserves the inner product with $|\phi \rangle$ is either $\mathds{1}$ or the reflection $\hat{O}_{|\phi \rangle} \equiv 2 |\phi \rangle  \langle \phi | - \mathds{1}$. The conserved quantities thus exactly  fix the forms of the orthogonal matrices that describe the collisions. 

\begin{itemize}
\item Collisions between the light ball and the wall, $\hat{O}_\textrm{wall}$. 

These collisions reverse the velocity of the light ball while leaving all other velocities unchanged $v_i \rightarrow (-1)^{\delta_{iw}} v_i$; alternatively we can say that they conserve the momentum of the large ball, 
\begin{equation}
\textrm{momentum of large ball} =  \sum_{i \neq w} v_i = \sqrt{d-1}  \langle \bar{s} | \theta \rangle \ . 
\end{equation}
Since the collision conserves $\langle \bar{s}  | \theta \rangle$,  symmetry tells us it must enact
\begin{equation}
\hat{O}_\textrm{wall} | {\theta} \rangle =  \Bigl( \mathds{1} - 2 |w \rangle \langle w | \Bigl) | {\theta} \rangle  =  \Bigl( 2 |\bar{s} \rangle \langle \bar{s} | - \mathds{1}  \Bigl) |\theta  \rangle  =  \Bigl(  |\bar{s} \rangle \langle \bar{s} | - |w \rangle \langle w |  \Bigl) |\theta  \rangle  \ . 
\end{equation}

\item Collisions between the heavy ball and the light ball, $\hat{O}_\textrm{ball}$. 

In the center-of-mass rest frame both velocities get reversed; alternatively we can say these collisions conserve the total momentum of the balls, 
\begin{equation}
\textrm{total momentum} = \sum_i v_i  = \sqrt{d}  \langle s | \theta \rangle.
 \end{equation} 
Since the collision conserves $\langle {s}  | \theta \rangle$, symmetry tells us it must enact
\begin{equation}
\hat{O}_\textrm{ball} |\theta \rangle =  \Bigl(\mathds{1} - 2 | \bar{w} \rangle \langle \bar{w} | \Bigl) | \theta \rangle = \Bigl(2 |s \rangle \langle s | - \mathds{1} \Bigl) |\theta \rangle  = \Bigl( |s \rangle \langle s | - | \bar{w} \rangle \langle \bar{w} | \Bigl) |\theta \rangle  .
\end{equation}
\end{itemize} 

  \begin{figure}[htbp] 
    \centering
    \includegraphics[width=\textwidth]{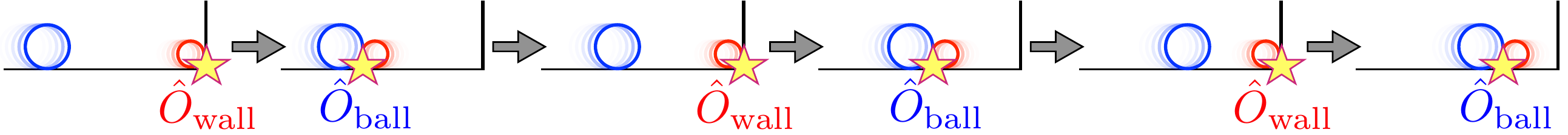} 
    \caption{The billiard bouncing back and forth between the wall and the large ball alternates $\color{red}{\hat{O}_\textrm{wall}}$ and $\color{blue}{\hat{O}_\textrm{ball}}$.}
    \label{fig-BackAndForth}
     \end{figure}

The small ball bounces back and forth between the wall ($\hat{O}_\textrm{wall}$) and the large ball ($\hat{O}_\textrm{ball}$). Each lap, the two reflections combine to make a rotation by $2 \bar{\theta}$, exactly as in Eq.~\ref{eq:multiplytorotation}
\begin{equation}
\hat{O}_\textrm{ball} \hat{O}_\textrm{wall} | \theta \rangle  =  | \theta + 2 \bar{\theta}  \rangle \  . 
\end{equation}

All that remains is to specify the initial state. Galperin's original $\pi$-counting plan \cite{Galperin} starts in $|\bar{s} \rangle$: the light ball begins at rest. To make the analogy with Grover's algorithm perfect, we should instead start in $| s \rangle$: the two balls begin with the same  velocity. With this minor tweak, the isomorphism between billiards and quantum search becomes exact.

  \begin{figure}[htbp] 
    \centering
     \includegraphics[height=1.9in]{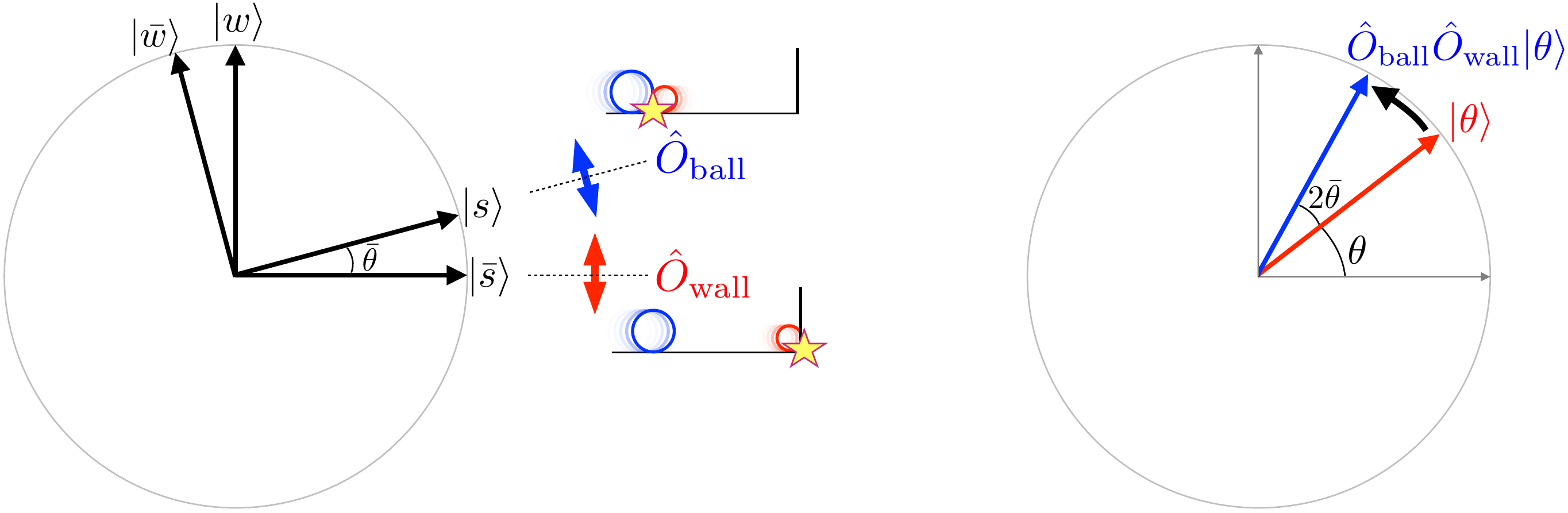}
    \caption{$\color{red}{\hat{O}_\textrm{wall}}$ reflects about $|\bar{s} \rangle$, and $\color{blue}{\hat{O}_\textrm{ball}}$ reflects about $| s \rangle$. In combination, ${\hat{O}_\textrm{ball}} {\hat{O}_\textrm{wall}}$ gives a rotation by $2 \bar{\theta}$.}
    \label{fig-statedictionary}
     \end{figure}

\section{Discussion} 
The isomorphism between Grover search and the bouncing of billiards is a duality between a discrete quantum system and a continuous classical system. The duality works for all values of $d$, though it is only for $d=100^N+1$ that Galperin's algorithm outputs the decimal digits of $\pi$. Let's examine some aspects of the duality in more detail. 
\begin{itemize}
\item Factor of Four. 

There is a factor of $4$ discrepancy between the collision-counting  Eq.~\ref{eq:numbercollisions} and the query-counting Eq.~\ref{eq:numbergrovercalls}. The factor of four is really two factors of two. 

One factor of 2 comes from how we count. The billiard problem counts every collision; by contrast Grover's problem treats $\hat{U}_s$ as free and only charges for applications of $\hat{U}_w$. 

The other factor of 2 comes from when we stop. Grover's algorithm stops when the heavy ball has transferred all its energy to the small ball to reach $| w \rangle = | \theta = \frac{\pi}{2} \rangle$; by contrast the billiard algorithm keeps going twice as long,  until the small ball has retransferred all its energy back to the heavy ball again to reach $- |\bar{s} \rangle = |\theta = {\pi} \rangle$. 

  \begin{figure}[htbp] 
    \centering
    \includegraphics[height=2in]{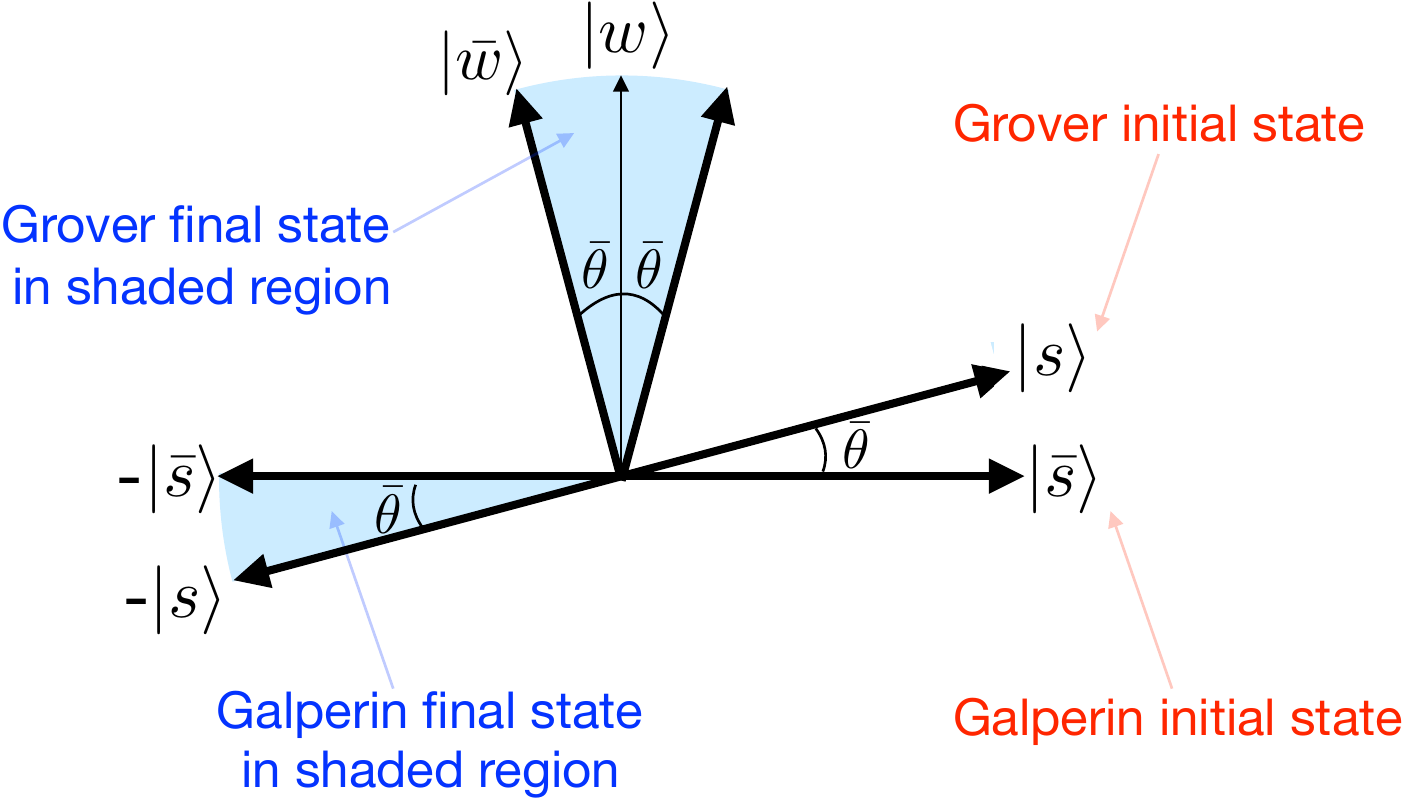}
    \caption{Galperin's protocol starts with the small ball stationary, $|\bar{s} \rangle$,  and ends once $v_{i \neq w} \leq v_w \leq 0$. Grover's algorithm starts in $|s \rangle$, with all the velocities equal, and ends at the closest approach to $|w \rangle$.}
    \label{fig-InitalAndFinal}
     \end{figure}

\item Surprising Squareroot. 

The crowd-pleaser in the collision-counting Eq.~\ref{eq:numbercollisions} is the ``$\pi$'', but even the square root is at first blush somewhat surprising. After all, the \emph{first} collision transfers only a single unit of kinetic energy, and there are $M$ units to be transferred in total, so you might  think it is going to take O($M$) collisions. The bouncing algorithm nevertheless gets the job done in O($\sqrt{M}$) steps because the momentum of the light ball grows linearly, so the kinetic energy grows quadratically.

The dual of this surprise occurs for Grover's algorithm. The \emph{first} iteration of $\hat{U}_s \hat{U}_w$ only increases the probability of measuring $|w \rangle$ by O($1/d$), but nevertheless only O($\sqrt{d}$) steps are needed to distinguish with near certainty.   Grover's algorithm gets the job done in O($\sqrt{d}$) steps because the amplitude $v_w$ grows linearly, so the probability grows quadratically.

\item One Hit Wonder. 

Grover search for one out of four items is unusually easy. With precisely one iteration of the algorithm, the search succeeds with 100\% probability. This is dual to the fact that when $M=3$, after a single pair of collisions the light ball is left exactly stationary. 
     \begin{figure}[htbp] 
    \centering
    \includegraphics[width=.9\textwidth]{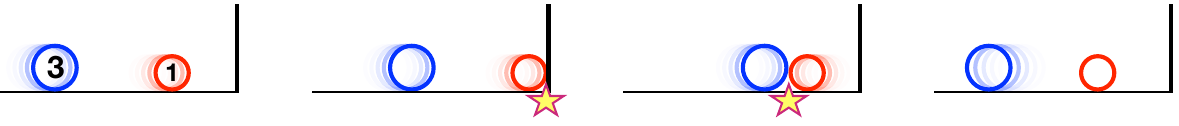} 
        \label{fig-OneHitWonder}
    \caption{One hit wonder. When ${{M}} = 3 $ and initially ${\color{blue}{v_\textrm{left} }} =  {\color{red}{v_\textrm{right}}}$,  a single pair of collisions leaves the small ball exactly stationary. Equivalently, Grover search for one of four items works exactly with a single iteration.}
 \end{figure}

\item Keeping it Real. 

Throughout the Grover process, the wavefunction stays real. The $v_i$ are always real, and operators $\hat{U}_w$ and $\hat{U}_s$ are not just unitary but orthogonal. Grover's algorithm thus shows you do not need complex numbers to get a quantum speed up. In the billiard problem, the reality of the wavefunction corresponds to the billiards moving only horizontally. 
\newpage

\item Numerous Needles. 

In the duality, the mass of the small ball is $1$ because there is $1$ correct answer (i.e. $1$ basis state whose sign is flipped by the oracle), and the mass of the heavy ball is $M = d-1$ because there are $d-1$ incorrect answers. 
If we were to adapt the Grover method to an oracle with $n$ correct answers ($n$ needles in the haystack) and $d-n$ incorrect answers, this would correspond to a mass $n$ for the small ball and mass $d-n$ for the heavy ball. 

 It is clear that if you double the mass of \emph{both} balls, the number of collisions does not change. Only the ratio matters. This is dual to the slightly less obvious fact that the runtime of the Grover task depends only on the ratio of the number of correct answers to the number of incorrect answers $\#_\textrm{queries} = \lfloor \frac{\pi}{4} \sqrt{(d-n)/n} \rfloor$.


\end{itemize}

Galperin's $\pi$-calculating plan displays a wanton disregard for engineering practicalities. \linebreak It requires that we overcome friction, overcome inelasticities, overcome the blurring effects of quantum mechanics, and then having overcome all these things it requires exceptional patience, because even pedestrian initial velocities provoke catastrophic corrections from special relativity. \linebreak 
Nevertheless, whatever the shortcomings of billiard balls as tools for calculating $\pi$, the results of this paper suggest a tool that is even worse. 
We might start with a qu$(100^N + 1)$it, and then step-by-step enact the quantum mirror of Galperin's method, mirroring each velocity with an amplitude, mirroring each billiard collision with a unitary, before ending with a painstaking tomographic reconstruction of the final state  and ushering in a new-if-pointless era of quantum arithmetic. 
It would not be easy,  it would not be useful, but it would be a picturesquely quixotic  way to seek $\pi$ in the $|\psi \rangle$.

\section*{Acknowledgements}
Thank you to Creon Levit and Michael Nielsen for feedback on a draft of this paper.

\end{document}